\documentclass[fleqn,usenatbib]{mnras}
\usepackage{newtxtext,newtxmath}
\usepackage[T1]{fontenc}
\usepackage{ae,aecompl}

\DeclareRobustCommand{\VAN}[3]{#2}
\let\VANthebibliography\thebibliography
\def\thebibliography{\DeclareRobustCommand{\VAN}[3]{##3}\VANthebibliography}

\usepackage{graphicx}	
\usepackage{amsmath}	

\usepackage{amssymb}	
\usepackage{multirow}
\usepackage{threeparttable}
\usepackage{hyperref}
\usepackage{ulem}

\newcommand{\ms}{\rm M_{\odot}}

\newcommand{\zl}{ z_{\rm l} }
\newcommand{\zs}{ z_{\rm s} }
\newcommand{\dif}{\,\mathrm{d}}

\usepackage{xspace}
\newcommand{\eagle}{\textsc{eagle}\xspace}

\title[Strong lens perturbations]{Galaxy-galaxy strong lens perturbations: line-of-sight haloes versus lens subhaloes}
\author[Q. He et al.]{Qiuhan He$^{1}$\thanks{E-mail: qiuhan.he@durham.ac.uk},
Ran Li$^{2,3}$\thanks{E-mail: ranl@bao.ac.cn},
Carlos S. Frenk$^{1}$,
James Nightingale$^{1, 4}$,
Shaun Cole$^{1}$,
\newauthor
Nicola C. Amorisco$^{1}$,
Richard Massey$^{1, 4}$,
Andrew Robertson$^{5}$,
Amy Etherington$^{1, 4}$,
\newauthor
Aristeidis Amvrosiadis$^{1}$,
Xiaoyue Cao$^{2, 3}$\\
$^{1}$Institute for Computational Cosmology, Department of Physics, Durham University, South Road, Durham DH1 3LE, UK\\
$^{2}$National Astronomical Observatories, Chinese Academy of Sciences, 20A Datun Road,Chaoyang District, Beijing 100012, China \\
$^{3}$School of Astronomy and Space Science, University of Chinese Academy of Sciences, Beijing 100049, China \\
$^{4}$Centre for Extragalactic Astronomy, Department of Physics, Durham University, South Rd, Durham, DH1 3LE, UK \\
$^{5}$Jet Propulsion Laboratory, California Institute of Technology, 4800 Oak Grove Drive, Pasadena, CA 91109, USA \\
}
\date{Accepted XXX. Received YYY; in original form ZZZ}

\begin{document}

\maketitle

\begin{abstract}
  We rederive the number density of intervening line-of-sight haloes
  relative to lens subhaloes in galaxy-galaxy strong lensing observations,
  where these perturbers can generate detectable image
  fluctuations. Previous studies have calculated the detection limit
  of a line-of-sight small-mass dark halo by comparing the lensing
  deflection angles it would cause, to those caused by a subhalo within the lens. However, this overly simplifies the difference in observational consequences between a subhalo and a line-of-sight halo. Furthermore, it does not take into account degeneracies between an extra subhalo and the uncertain properties of the main lens. More in keeping with analyses of
  real-world observations, we regard a line-of-sight halo as
  detectable only if adding it to a smooth model generates a statistically significant improvement in the reconstructed image. We find that the
  number density of detectable line-of-sight perturbers has been
  overestimated by as much as a factor of two in the previous
  literature. For typical lensing geometries and configurations, very deep imaging is sensitive to twice as many line-of-sight perturbers as subhaloes, but moderate depth imaging is sensitive to only slightly more line-of-sight perturbers than subhaloes. 
\end{abstract}
\section{Introduction}
The most fundamental prediction of the cold dark matter (CDM)
cosmological model is the existence of a large population of low mass
dark matter haloes
\citep{Green2005,Diemand2007,Springel2008,Frenk2012,Wang2020}. This
feature can be used to test the CDM model rigorously or to distinguish
it from models with alternative types of dark matter. For example, the
warm dark matter (WDM) model predicts a cutoff in the power spectrum
of initial density perturbations, induced by particle free streaming,
which translates into a cutoff in the halo mass function at a mass
scale that depends on the WDM particle mass
\citep{Colin2000,Lovell2012,Schneider2012,Bose2016}.

Recent observations have revealed a line in the X-ray spectra of
galaxies and galaxy clusters at 3.5~keV that could result from decay
of a WDM particle such as a 7~keV sterile neutrino
(\citealt{Boyarsky2014, Bulbul2014}, but see
\citealt{Riemer-Sorensen2016}). In this case, the halo mass function
today would exhibit a sharp cutoff at a mass,
$m_{\rm 200} \lesssim 10^8$ $\ms$. Thus, this model would be
conclusively ruled out if one could demonstrate the existence of a
population of haloes below this mass scale.  Conversely, the CDM model
would be conclusively ruled out if such a population were not found.
Haloes that are of mass $m_{\rm 200} \lesssim 3 \times 10^8$ $\ms$ today were
never able to make stars and therefore remain completely dark \citep[see][and
references therein]{Benitez-Llambay2020,Sawala2016}. Such haloes
cannot therefore be detected by conventional means but they can, in
principle, be detected through their gravitational lensing effects.

Strong lensing systems that exhibit giant arcs or Einstein rings can
appear measurably perturbed if any of the light from the source passes
sufficiently close to a small dark halo
\citep{Koopmans2005,Vegetti2009a,Vegetti2009b,Vegetti2012,Hezaveh2016,Li2016}. This
is a difficult measurement but applying sophisticated data analysis
and modelling tools to high resolution imagery, it is possible to
detect small haloes projected near the Einstein radius of the lens and
infer their mass \citep[e.g.][]{Koopmans2005, Vegetti2009a}.  The mass detection limit for dark matter haloes
depends on the resolution of the image. With Hubble Space Telescope
(HST) imagery, \citet{Vegetti2010} discovered a dark perturber of
mass, $3.51 \pm 0.15 \times 10^9$~M$_{\odot}$, which they interpret
as a subhalo in the strong lens system SDSS J0946+1006 (the mass here
refers to that of a truncated pseudo-Jaffe density profile).  Another
dark object of mass, $1.9 \pm 0.1 \times$ $10^8$M$_{\odot}$, was found
in a lens galaxy at redshift $z=0.88$ from even higher resolution
imaging using adaptive optics at the Keck telescope
\citep{Vegetti2012}.

To constrain the nature of the dark matter with this technique it is,
of course, necessary to know the expected number of lensing
perturbers in CDM and other models of interest. The perturbers can
be either subhaloes of the main lens or ``field'' haloes that are not
part of the main lens but appear projected near its Einstein
radius. In what follows, we will refer to the latter as
``line-of-sight'' or ``intervening'' haloes or perturbers.

To count the expected number of subhaloes and intervening haloes
consistently, \citet[][hereafter Li17]{Li2017} derived an effective mass,
$M_{\rm eff}$($M_{\rm los}$), for a line-of-sight halo of mass,
$M_{\rm los}$. This is determined by fitting a lensing image perturbed
by an NFW halo at the lens redshift to the image  perturbed by a
line-of-sight halo of mass, $M_{\rm los}$, at redshift, $z_{\rm
  los}$. In other words, if a subhalo of mass,
$M_{\rm eff}$($M_{\rm los}$), can be detected, a line-of-sight halo of
mass, $M_{\rm los}$, should also be detected. In this way, one can
calculate the effective mass function of all
perturbers. Li17 showed that, for CDM, the number of
detectable line-of-sight perturbers is 3-4 times larger than the
number of subhalo perturbers.

A similar analysis was performed by \citep[][hereafter D18]{Despali2018}, who derived a
fitting formula for $M_{\rm eff}$ by fitting the deflection angle of a
lensing system containing a line-of-sight perturber to that of a
system containing a subhalo. The analysis was performed for lenses
with different image and redshift configurations. In agreement with
the results of Li17, they found that the contribution from
CDM line-of-sight haloes is about 3 times that from subhaloes for
lenses with $\zl=0.2$ and $\zs=1.0$, and about 10 times that from
subhaloes for lenses with $\zl=0.5$ and $\zs=2.0$. Besides, by statistically studying low mass haloes' perturbation, \citet{Cagan2020} also show that line-of-sight perturbers tend to dominate the signal for systems with a source at redshift higher than 0.5.

The preliminary conclusion that the lensing distortions are dominated
by line-of-sight haloes is encouraging because it greatly simplifies the
theoretical analysis. Unlike for subhaloes, whose mass function is
affected by environmental effects, calculating the mass function of
dark central haloes in the mass range of interest - below the
threshold for star formation - is straightforward since these haloes
have never been affected by baryons. Thus, a standard calculation of
the mass function based on dark-matter-only simulations
\citep{Frenk1988} 
gives very precise results \citep[see][for a recent review]{Zavala2019}.

By contrast, the mass function of subhaloes is determined by a number
of processes, such as tidal stripping or tidal shocking, that alter 
the mass distribution and can destroy the subhalo. To calculate these
processes requires modelling the host galaxy in detail, including its
baryonic components. This is, of course, a much more complicated
problem than simply following the evolution of dark matter 
haloes. Significant advances, however, have been achieved in recent
years with a new generation of cosmological hydrodynamics simulations
that can produce realistic galaxy populations
\citep[e.g.][]{Vogelsberger2014,Schaye2015}. Here, we will make
extensive use of the high-resolution hydrodynamics simulation of
\cite{Richings2021} of a galaxy cluster and its environment which
includes the relevant baryon physics processes. 
 

Li17 and D18 both made an important assumption:
that the perturbation induced by a subhalo can always be well fitted
by the perturbation induced by a line-of-sight halo with an NFW
profile. This assumption, however, may not be exactly true because the
deflection angles produced by a line-of-sight halo can be very
different from that of a subhalo in certain redshift
ranges. Furthermore, these earlier studies did not carry out complete
modelling of the lensing process, for example, assuming realistic
noise levels, and this may further bias the results.

In this paper, we revisit the importance of the contribution of
line-of-sight perturbers by modelling a set of realistic strong lensing mock images. In earlier studies, the comparison of deflection angles was used to decide whether a perturber is detectable or not, through the concept of an ``effective mass''. Here we derive the detectable mass threshold for
line-of-sight perturbers by directly evaluating the difference in
log-likelihood between a model with a perturber and a model without a
perturber, using a state-of-the-art strong lens modelling
pipeline \textbf{\sc
  PyAutoLens}\footnote{https://github.com/Jammy2211/PyAutoLens}
\citep{Nightingale2018, pyautolens}. The new threshold is now directly obtained from modelling image fluxes and thus it is more straightforward and robust, where it takes into account factors from flux modelling processes previously not considered, like the degeneracy between the perturber and the macro model. We also investigate the dependence of the
relative contribution of the two types of object on the redshift and the
S/N ratio of the observations.

An independent study of the sensitivity function using \texttt{PyAutoLens} 
is provided by \citet{Amorisco2022}. This work reassuringly reaches 
the same conclusion as us on the dependence of the sensitivity function on the 
redshift of the perturbing halo, despite using a different approach to calculate
the sensitivity function and mock strong lens datasets with different properties. 
Looking at only line-of-sight haloes, this study 
highlights the impact that the intrinsic scatter in halo concentrations has 
on the sensitivity function, and shows that the dependency of the concentration--mass 
relation on the dark matter model improves strong lensing as a probe of dark matter.
In this work, we explicitly include subhalos within the lens galaxy (in addition to 
line-of-sight haloes), accounting for their different mass function and density profiles
due to baryonic physics. Thus, we set out to answer whether the line-of-sight or 
lens galaxy subhalos dominate the constraints on DM.

The structure of the paper is as follows. In Section~\ref{sec:method}, we
describe how we construct mock lenses, how we compute the sensitivity
map and the method we use to translate sensitivity maps into constraints on
the halo mass function. In Section~\ref{sec:results}, we present our
results and in Section~\ref{sec:conclusion}, we summarize our
conclusions. Throughout the paper we adopt the Planck cosmological
parameters \citep{Planck2016}:
$\rm H_0 = 67.7 \; km\,s^{-1}\,Mpc^{-1}$, $\rm \Omega_m = 0.307$ and
$\rm \Omega_\Lambda = 0.693$.

\section{Method}\label{sec:method}
\subsection{Mock Lenses}\label{sec:mock_lens}

We construct five sets of mock lenses, including examples with
different image configurations, redshifts, noise levels and angular
resolution. For simplicity we set the density distribution of all primary
lenses to be singular isothermal ellipsoids (SIE),
\begin{equation}
    \Sigma(x, y) = \frac{{\rm c}^2 }{8 \mathrm{\uppi G}}\frac{D_{\rm A} \left(0,\ z_{\rm s}\right)}{D_{\rm A} \left( z_{\rm l},\ z_{\rm s}\right) D_{\rm A}\left(0,\ z_{\rm l}\right)} \frac{R_{\rm E}}{\sqrt{x^2 q + y^2/q}},
\end{equation}
where $R_{\rm E}$ and $q$ are the Einstein radius and axis ratio of
the lens galaxy; $D_{\rm A} \left( z_1,\ z_2\right)$ is the angular
diameter distance between redshifts, $z_1$ and $z_2$. The lens and
source redshifts are marked as, $z_{\rm l}$ and $z_{\rm s}$,
respectively. We do not add external shear in the mock lenses. However, when
modelling the lens we do include the external shear as part of our mass
model \citep{Witt1997}.

To simulate the source galaxies, we assume a ``cored'' Sersic density
profile:
\begin{equation}
    I(r) = I^{'}{\rm exp}\left[- b_n \left(\frac{r^\alpha + r^\alpha_{\rm c}}{r_{\rm e}^\alpha}\right)^{1/(n\alpha)}\right],
\end{equation}
where $I^{'}$ is the scale intensity, $r_{\rm e}$ the effective
radius, $n$ the Sersic index and $b_{\rm n}$ a coefficient related to
the Sersic index (see Eq.~A7 of \citealt{Trujillo2004}). Compared to
the standard Sersic profile, the core model introduces two additional
parameters, $r_{\rm c}$, which describes the core size, and $\alpha$,
which controls how fast the profile approaches constant surface
brightness inwards. Throughout our tests we fix $\alpha=2.0$ and
$r_{\rm c}=0.01$\arcsec. The small core in our model helps to remove
potential numerical inaccuracies induced by the cuspy nature of the
regular Sersic profile.

The fiducial mock image setup has a nearly complete Einstein ring with
the lens galaxy at redshift, $z= 0.5$, and the source galaxy at
$z=1$. The emission of the lens galaxy is omitted in this work. The mock image has 
similar angular resolution to HST imaging, where the pixel size is $0.05\arcsec$ and a Gaussian Point Spread 
Function (PSF) is assumed where $\sigma = 0.05\arcsec$ (FWHM of $\sim 0.118\arcsec$). 
For the noise level, we try to set it to be similar to the best cases in the SLACS
sample \citep{Bolton2006}, where the maximum S/N in the image pixels is around 40 for 
a 2000s exposure. We adopt a background sky noise level of 0.1~e$^{-}\,\mathrm{pix}^{\rm -1}\,\mathrm{s}^{\rm -1}$, 
which is estimated from HST images of SLACS lenses. To add noise to each mock image, the background sky  is added to the lensed 
source image, the data is converted to units of counts and Poisson noise values are drawn and added to every pixel. The source 
intensity is adjusted to make the maximum pixel S/N be $\sim$ 40 and the data is then converted back to e$^-{\rm s}^{\rm -1}$. The 
lens and source parameters of the fiducial setting is shown in the third column of Table~\ref{tab:parameters}.

\begin{table}
\begin{center}
\begin{tabular}{c c c c}
\hline
        &  & Einstein Ring & Quad \\
\hline
        \multirow{2}{*}{Lens} & (x, y) [(\arcsec, \arcsec)] & (0.0, 0.0) & (0.0, 0.0) \\
        & $R_{\rm E}$ [\arcsec] & 1.5 & 1.5 \\
        & q & 0.95 & 0.65 \\
        & $\theta$ [$^\circ$] & 30 & 30 \\
\hline
      \multirow{2}{*}{External Shear} & magnitude & 0.0 & 0.0 \\
      & $\theta$ [$^\circ$] & 0.0 & 0.0 \\
\hline
        \multirow{2}{*}{Source} & (x, y) [(\arcsec, \arcsec)] & (0.1, 0.1) & (-0.05, 0.1) \\
        & $r_{\rm e}$ [\arcsec] & 0.2 & 0.2 \\
        & q & 0.52 & 0.7 \\
        & $\theta$ [$^\circ$] & 30 & 30 \\
        & $I^{'}$ [e$^{-}$~pix$^{-1}$~s$^{-1}$] & 2.2 & 2.3 \\
        & $n$ & 2.5 & 2.0 \\
        & $r_{\rm c}$ [\arcsec] & 0.01 & 0.01 \\
        & $\alpha$ & 2.0 & 2.0 \\
\hline
        
\end{tabular}
\caption{Parameters of the lens and source galaxies for the Einstein Ring and Quad images in our mock simulations. }\label{tab:parameters} 
\end{center}
\end{table}

\begin{figure*}
    \includegraphics[width=0.8\textwidth]{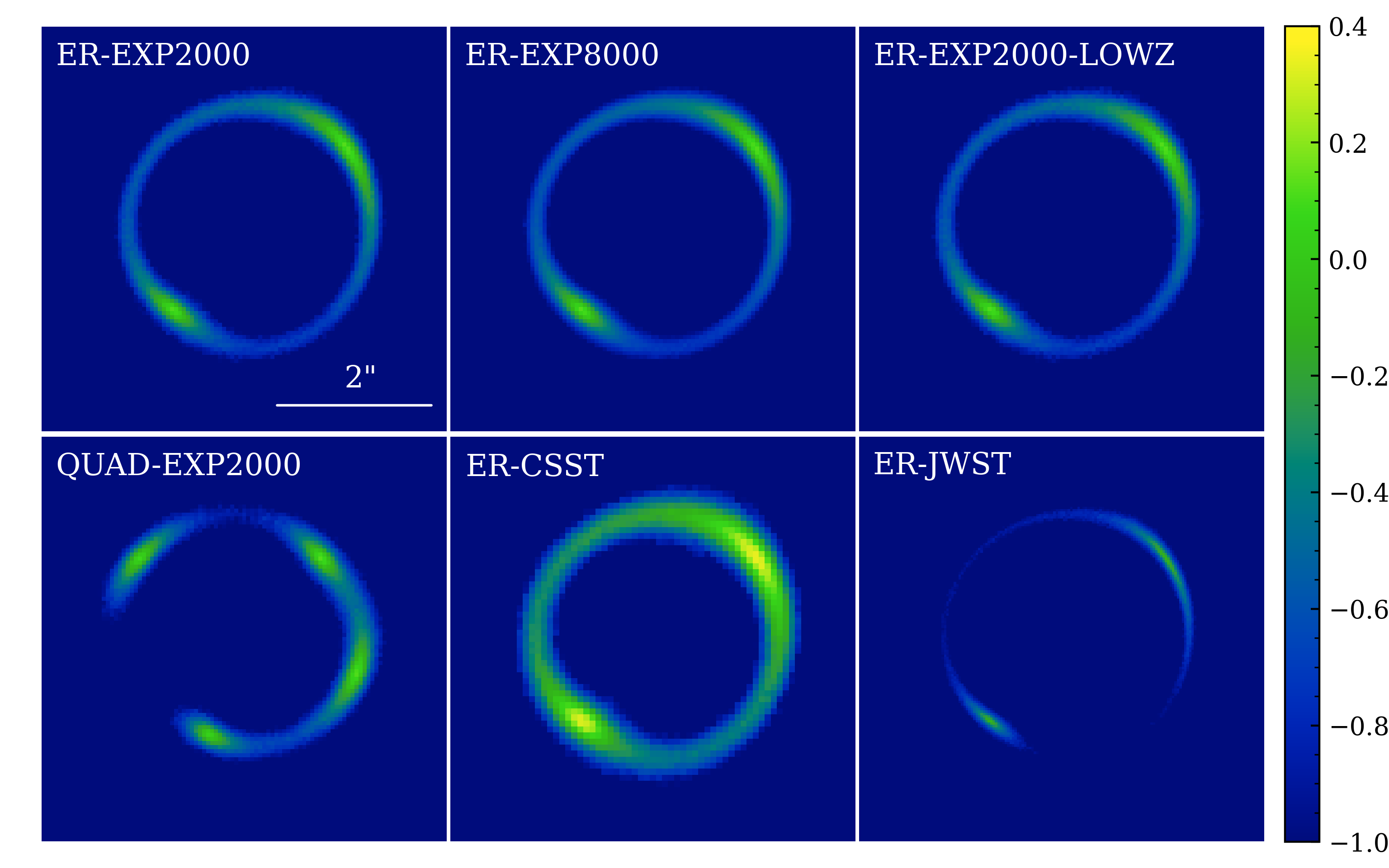}
    \caption{Mock observations of the galaxy-galaxy strong lens systems that we investigate. The name of each setting is shown on the top left of each panel. "ER-EXP2000" is our fiducial mock image. "ER-EXP8000" has a 8000s exposure. "ER-EXP2000-LOWZ" has a main lens located at z = 0.22. "QUAD-EXP2000" is an image with quadruple arcs. "ER-CSST" uses the resolution of CSST, which has a pixel size of $0.075\arcsec$ and a PSF sigma of $0.08\arcsec$ (FWHM of $0.188\arcsec$). "ER-JWST" uses the resolution of JWST, which has a pixel size of $0.03\arcsec$ and a PSF sigma of $0.013\arcsec$ (FWHM of $0.03\arcsec$). Parameters of the lens and source parameters for the Einstein Ring and Quad images are listed in Table~\ref{tab:parameters}. The key features of the systems are listed in Table~\ref{tab:mock}. The images are shown in log$_{10}$ scale, and the unit of the color bar is e$^{-}$~pix$^{-1}$~s$^{-1}$. }
    \label{fig:mocks}
\end{figure*}

Based on this fiducial setup, we also change the appropriate
parameters to explore the effects of different image configurations,
lens galaxy redshifts, noise levels and angular resolution. We summarize our mock images (without adding any low mass haloes) in Fig.~\ref{fig:mocks}, where each setting's name is labelled in the upper left of each panel. The details of each setting are as follows:
\begin{itemize}
    \item ER-EXP2000 --- This is our fiducial setting, a nearly complete Einstein ring with a radius of $1.5\arcsec$,  corresponding to an Einstein mass of $6.4\times10^{11}$~M$_\odot$ within 9.4~kpc.
    \item ER-EXP8000 --- It has the same setting as our fiducial mock except the exposure time increases to 8000s, of which the maximum pixel S/N is $\sim$ 80. 
    \item ER-EXP2000-LOWZ --- The main lens in this case is located at redshift,  $z=0.22$, while the source is still at $z = 1$. When changing the lens redshift, we keep its Einstein radius (in arcsec) and noise level unchanged. At this redshift, a 1.5\arcsec\ Einstein radius corresponds to an Einstein mass of $1.3\times10^{11}$~M$_\odot$ within 5.5~kpc. 
    \item QUAD-EXP2000 --- In this setting, we simulate an image with quadruple arcs with the same noise level and angular resolution as the fiducial case. The lens and source parameters of this configuration are listed in the fourth column of Table~\ref{tab:parameters}.
    \item ER-CSST --- We also simulate a configuration with lens and source properties as in the fiducial case but with the same image resolution as the China Space Station Telescope (CSST). The CSST resolution is slightly worse than for the HST, with a pixel size of $0.075\arcsec$ and a PSF $\sigma$ of $0.08\arcsec$ (FWHM of $\sim$ $0.188\arcsec$). We note that the hardware design of the CSST is not fully determined yet, so for our purpose here, which is to study  purely resolution effects, we assume it has similar noise conditions as the HST except for the resolution. Since the pixel area is 2.25 times larger, the background sky noise is also increased to 0.225~e$^{-}\,\mathrm{pix}^{\rm -1}\,\mathrm{s}^{\rm -1}$. The image simulated here may therefore be treated as a pixel binned version of the fiducial mock image using a larger pixel and PSF size of the CSST.
    \item ER-JWST --- The image resolution of this case is close to that  of James Webb Space Telescope (JWST) images, which have a pixel size of $0.03\arcsec$ and a PSF $\sigma$ of $0.013\arcsec$ (FWHM of $\sim$ $0.03\arcsec$). Note that now the pixel areas are 0.36 times those of the previous case;  the noise is also changed consistently to be 0.036~e$^{-}\,\mathrm{pix}^{\rm -1}\,\mathrm{s}^{\rm -1}$. 
\end{itemize}
In Table~\ref{tab:mock} we summarize the key features of the five mock
settings (the quantities of the last three columns are defined in our results part, Sec.~\ref{sec:results}).

\begin{table*}
\begin{center}
\begin{threeparttable}
\begin{tabular}{c c c c c c c c c c}
\hline
        Label & Configuration & $\zl$ & $\zs$ & Exposure time (s) & $\sigma_{\rm PSF}$ & Pixel size &  $N_{\rm los}$ & $N_{\rm sub}$ & $N_{\rm los}/N_{\rm sub}$ \\
\hline 
        ER-EXP2000 & Einstein Ring & 0.5  & 1.0  & 2000 & 0.05 & 0.05 & 0.85 & 0.66 & 1.29 \\
\hline 
        ER-EXP8000 & Einstein Ring & 0.5  & 1.0  & 8000 & 0.05 & 0.05 & 2.50 & 1.24 & 2.02 \\
\hline 
        ER-EXP2000-LOWZ & Einstein Ring & 0.22  & 1.0  & 2000 & 0.05 & 0.05 & 0.39 & 0.38 & 1.03 \\
\hline 
        QUAD-EXP2000 & Quad & 0.5  & 1.0 & 2000 & 0.05 & 0.05  & 0.64 & 0.51 & 1.25 \\
\hline
        ER-CSST & Einstein Ring & 0.5  & 1.0  & N.A.\tnote{1} & 0.08 & 0.075 & 0.72 & 0.60 & 1.20\\
\hline
        ER-JWST & Einstein Ring & 0.5 & 1.0 & N.A.\tnote{1} & 0.03 & 0.03 & 1.06 & 0.75 & 1.41 \\
\hline

\end{tabular}
\begin{tablenotes}
\item[1] For our purpose to investigate how image resolution affect our results, we set them to have equivalent depth of observation as our fiducial setting.
\end{tablenotes}
\end{threeparttable}
\caption{Column 1 shows the label of six mocks. Column 2-6 show the settings of lens configuration, including lens reshift, source redshift, exposure time, PSF size and pixel size. Column 7-8 show the expected detection of line-of-sight haloes, the subhaloes per system, and Column 9 shows the their ratio. Parameters of the Einstein Ring and Quad lens configurations are listed in Table~\ref{tab:parameters}.}\label{tab:mock}
\end{center}
\end{table*}

We perturb the images of the mock lenses with two types of objects.  One are
small-mass dark matter haloes along the line-of-sight and the other
are subhaloes within the host halo of the lens galaxy. We model the
line-of-sight haloes with spherical NFW profiles using the
mass-concentration relation given by \citet{Ludlow2016}, which has
been shown to match the simulation data very well at the low mass
end \citep{Wang2020}.


For subhaloes associated with the lens, their mass profile in the
outer parts is modified by environmental effects such as tidal
stripping. N-body simulations have shown that the density of these
subhaloes drops dramatically beyond a truncation
radius \citep[e.g.][]{Gao2004}. In this paper we simulate the subhalo
profile using a truncated NFW (tNFW) profile \citep[see Eq.~A.26$\sim$A.33 of][]{Baltz2009},
\begin{equation}
    \rho\left(r\right) = \frac{m_0}{4{\rm \pi}}\frac{1}{r\left(r + r_{\rm s}\right)^2}\left(\frac{r_{\rm t}^2}{r_{\rm t}^2 + r^2}\right)^2,
\end{equation}
where $m_0$ is the scale mass, $r_{\rm s}$ is the scale radius and
$r_{\rm t}$ is the truncation radius. The total mass of the subhalo,
$m_{\rm tot}$, can be written as \citep[Eq.~A.29 of][]{Baltz2009}, 
\begin{equation}
    m_{\rm tot} = \frac{m_{0} \tau^2}{2\left(\tau^2 + 1\right)^3}\left[2\tau^2\left(\tau^2-3\right){\rm ln}\tau - \left(3\tau^2 - 1\right)\left(\tau^2 + 1 - \tau{\rm \pi}\right)\right],
    \label{eq:mtot_tnfw}
\end{equation}
where $\tau \equiv r_{\rm t} / r_{\rm s}$.

We derive the mean relation between $r_s$, $r_t$ and $m_{\rm tot}$
using the high-resolution hydrodynamical simulation by
\citet{Richings2021}. This is a zoom, high resolution  resimulation of a halo with $m_{200}$ (the mass within $r_{200}$, the radius where the enclosed density is 200 times the critical density of the Universe) of $10^{13.1}$~M$_{\odot}$ at $z = 0.18$ selected from the \eagle simulation volume
\citep{Schaye2015} and resimulated with 17 times better gas mass
resolution than \eagle, $m_g=1.8\times 10^5$~M$_{\odot}$, and about 100
times better dark matter mass resolution than \eagle,
$m_{\rm DM}=8.3\times 10^4$~M$_{\odot}$. Such high resolution allows us to
resolve the internal structure of subhaloes more massive than 10$^8$
M$_\odot$. Haloes were identified using the friends-of-friends algorithm
\citep{Davis1985} and subhaloes using the \textbf{\sc SUBFIND}
algorithm \citep{Springel2001}; the density profiles of subhaloes in
the mass range $10^{8}\sim10^{11}$ M$_\odot$ were fit with the tNFW
formula to derive the values of $r_{\rm s}$, $r_{\rm t}$ and $m_{\rm tot}$.

Since in actual observations, only subhaloes around the Einstein
radius matter, we only select subhaloes whose projected positions fall
in an annular region between $0.5\arcsec$ and $3.0\arcsec$ from the lens centre to derive the relations
between $r_s$, $r_t$ and $m_{\rm tot}$. To improve the statistics, we
rotated the simulated halo 10000 times at random. For the mock lenses at
$z=0.22$ and $z=0.5$ we make use of the snapshots at $z=0.183$ and
$z=0.503$ respectively. To make sure the derived relations are not
dominated by any particular subhalo, e.g. by one very close to the
Einstein radius in 3D, we carry out bootstrap tests whereby we repeat
the same procedure 200 times and each time we derive the linear
relations from a random re-sample of all subhaloes. The linear relation is computed by minimizing a $\chi^2$ defined as:
\begin{equation}
    \chi^2 = \sum_{\rm i=1}^{\rm N} W_{\rm i} \left(y_{\rm i} - \left(m \times x_{\rm i} + c\right)\right)^2,
\end{equation}
where $m$, $c$ are the linear relation slope and intercept, $\left(x_{\rm i}, y_{\rm i}\right)$ is the coordinate of the $i$-th data point and $W_{i}$ is the number of times the $i$-th data point is repeated in the sample.
We then
take the median linear relations from the 200 tests to model the
subhaloes.

\begin{figure}
        \centering
        \includegraphics[width=0.5\textwidth]{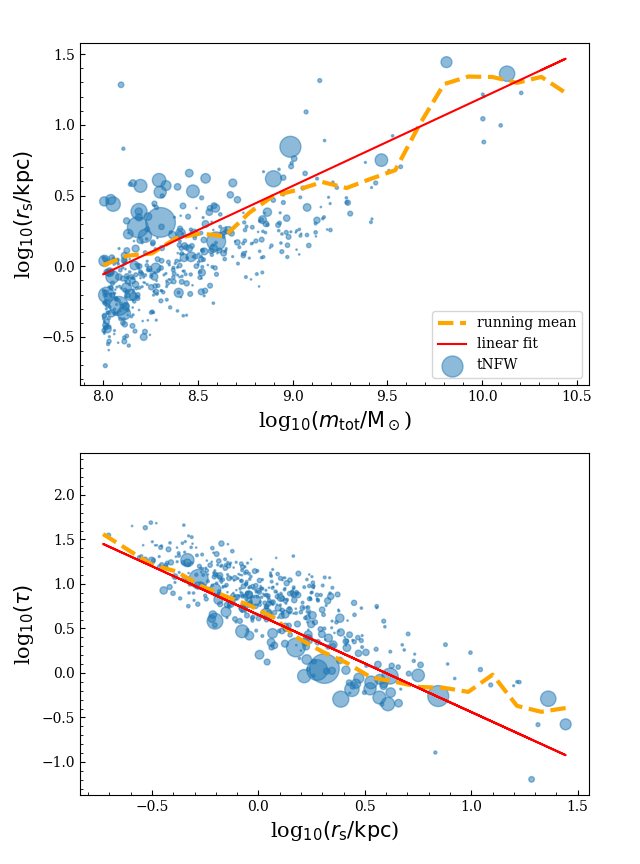}
        \caption{Relations between $m_{\rm tot}$, $r_{\rm s}$ and
          $\tau$ for subhaloes in an annulus between $1.0\arcsec$ and $3.0\arcsec$
          encompassing the Einstein radius, at snapshot $z= 0.183$. The
          relations were obtained by rotating the lensing galaxy and its subhaloes 10000 times
          and selecting those subhaloes that fall in the region of interest in
          projection. The area of each blue point represents how many
          times a subhalo falls in this region. The dashed orange and solid red lines show the
          running means and
          best fit linear relations for the data taking account of the weight of each data point.}
        \label{fig:mtot_rs_tau}
\end{figure}

Taking the snapshot at $z = 0.183$ as an example, in
Fig.~\ref{fig:mtot_rs_tau}, we plot the values of $m_{\rm tot}$,
$r_{\rm s}$ and $\tau \left(r_{\rm t}/r_{\rm s}\right)$ for subhaloes
as blue circles. The area of each circle reflects how many times each
subhalo fell on the annular region around the Einstein radius, which is taken as a weight for the point when deriving the linear relation. For
this lens system we find,
\begin{equation}
\begin{split}
    & {\rm log}_{10}\left(\frac{r_{\rm s}}{\rm kpc}\right) = \left(0.60\pm0.09\right){\rm log}_{10}\left(\frac{m_{\rm tot}}{{\rm M}_{\odot}}\right) - \left(4.85\pm0.77\right) \\
    & {\rm log}_{10}\left(\tau\right) = \left(-1.21\pm0.10\right){\rm
      log}_{10}\left(\frac{r_{\rm s}}{\rm kpc}\right) +
      \left(0.59\pm0.04\right). 
\end{split}
\end{equation}

For the lens at $z=0.5$ we find,

\begin{equation}
\begin{split}
    & {\rm log}_{10}\left(\frac{r_{\rm s}}{\rm kpc}\right) = \left(0.49\pm0.06\right){\rm log}_{10}\left(\frac{m_{\rm tot}}{{\rm M}_{\odot}}\right) - \left(3.90\pm0.54\right) \\
    & {\rm log}_{10}\left(\tau\right) = \left(-1.21\pm0.09\right){\rm log}_{10}\left(\frac{r_{\rm s}}{\rm kpc}\right) + \left(0.55\pm0.03\right),
\end{split}
\end{equation}
where the errors are the 1$\sigma$ scatter determined from the
bootstrap resampling. 

\begin{figure}
        \centering
        \includegraphics[width=0.5\textwidth]{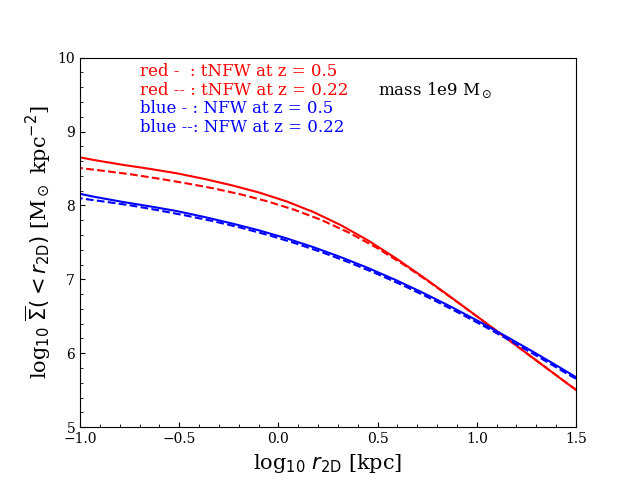}
        \caption{Interior mean surface density profiles of tNFW and NFW at $z = 0.22$ (dashed lines) and $z = 0.5$ (solid lines). Red lines are the tNFW profiles and blue lines show the NFW profiles. All the haloes plotted here have mass of 10$^{9}$~M$_\odot$. For tNFW, the mass refers to the total mass, while for NFW, the mass refers to $m_{200}$.}
        \label{fig:profile_NFW_tNFW}
\end{figure}

In Fig.~\ref{fig:profile_NFW_tNFW}, we show the interior mean surface density profiles of tNFW
subhaloes of mass $m_{\rm tot}=10^9$~M$_\odot$ for two different lens
redshifts, and the same profiles of central haloes of mass
$m_{200}=10^9$~M$_\odot$ for two different lens redshifts. The profiles of the tNFW model were derived
from the mean linear relations between $m_{\rm tot}$, $r_s$, and $r_t$
described above. Clearly, the tNFW profiles are more compact and have
higher amplitude than the NFW profiles of the same mass.

\subsection{Sensitivity Mapping}

A process called sensitivity mapping is performed to quantify the detectability of a perturbing halo that is nearby a strongly lensed source. One begins by modeling a strong lens dataset to infer an accurate model for the lens's mass and source's light \citep{Nightingale2019}. Using this model, one can then simulate a new realization of the strong lens which includes a dark matter perturber at a given $(x, y)$ position in the image-plane and with an input mass and redshift. This simulated dataset assumes the same image resolution and PSF of the true dataset and also has consistent signal-to-noise properties. 

The mock dataset is now fitted with two lens models: (i) a lens mass model which does not include a dark matter 
perturber and; (ii) a lens mass model which does. By comparing a goodness-of-fit measure of each model-fit 
(e.g. the maximum log likelihood value) one therefore quantifies how sensitive the lens dataset is to a dark matter
perturber, given its input location and mass. If the lens model including the perturber has a much improved goodness-of-fit compared to the model which does not, the perturber was necessary to fit the data accurately, indicating that the strong lens data is sensitive to perturbers at the location and with that mass. If the goodness-of-fits
are comparable, the perturber does not improve the lens model and therefore it is too far from the lensed source or too low mass to be detectable. 

It is necessary to perform two full fits to each mock dataset, to infer the maximum log likelihood of each model, for two reasons. First, the image fluxes we are fitting have noise and thus the maximum likelihood model may not be the true input model  \citep[although see][]{Amorisco2022}. Second, due to the existence of a small perturber in the mock data and the possible degeneracy between it and the main lens mass, when fitting a model without a small perturber, the maximum likelihood model is offset with respect to the true input and can only be found via a full fit.

\begin{figure}
        \centering
        \includegraphics[width=0.5\textwidth]{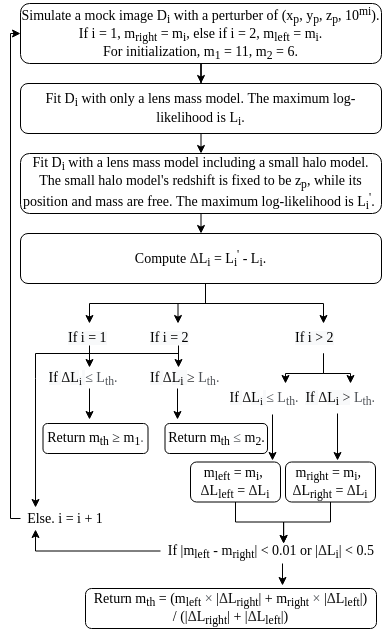}
        \caption{The binary search procedure for $m_{\rm th}$ at ($x_{\rm p}$, $y_{\rm p}$, $z_{\rm p}$). $L_{\rm th}$ is the detection threshold and throught this paper, we take it to be 10.}
        \label{fig:procedure}
\end{figure}

By repeating this process on a grid of perturber ($x$,$y$) location, mass and redshift one produces a sensitivity map. In this work, we assume a grid of 27 steps in the $y$ and $x$ directions, and a grid of 25 steps in redshift between $z$ = 0.02 and $z$ = 0.98. For a given object, its angular size decreases with redshift, so the angular size of the sensitive region also decreases with redshift. To save computational resources, we decrease the angular size of the explored region as the  redshift increases. Furthermore, for the same reason, rather than exploring a grid of mass values, we look for the lowest value of the perturber's mass that corresponds to the goodness-of-fits threshold of detection by a binary search algorithm (in ${\rm log}_{10}$ scale). The mass boundaries for the binary search are 10$^6$~M$_\odot$ and 10$^{11}$~M$_\odot$ and the stop criterion for the iteration is that $\left|\Delta {\rm log}_{10}\left(m\right)\right| < 0.01$. For every grid cell, we simulate a new strong lens dataset and fit it with the two lens models described above using the nested sampling algorithm \texttt{dynesty} \citep{dynesty}. For efficiency, we use tight priors on every model-fit that exploit our knowledge of what values of lens mass model and source model were used to when simulating the data. This could negatively impact \texttt{dynesty}'s estimate of the Bayesian evidence, therefore we opt to simply compare maximum log likelihood values when producing a sensitivity map. At the end, the ``sensitivity map'' is a grid of ($x_{\rm p}$, $y_{\rm p}$, $z_{\rm p}$, $m_{\rm th}$), which means that a perturber at ($x_{\rm p}$, $y_{\rm p}$, $z_{\rm p}$) is detectable when it has a mass over the threshold mass, $m_{\rm th}$. Please note when fitting the mock image with a lens mass model including a pertuber model, the perturber model's redshift is fixed to be $z_{\rm p}$ while its position and mass are free (the concentration follows \citet{Ludlow2016} which is a function of $m$ and $z_{\rm p}$.). Fig.~\ref{fig:procedure} summarizes the procedure of computing $m_{\rm th}$ for at ($x_{\rm p}$, $y_{\rm p}$, $z_{\rm p}$). Our method is conceptually analogous to that of \citet{Amorisco2022}, albeit there are differences in the fitting algorithm used.

\subsection{Detection threshold}

Previous studies \citep{Li2016, Despali2018} have derived the
detection threshold of a line-of-sight perturbing halo by directly
modelling the lensing effect of a subhalo, i.e. a perturber at the redshift of the lens. If a line-of-sight halo of
mass, $m$, best fits the lensing effect of a subhalo of mass,
$m_{\rm sub}$, then this is defined as the effective mass of the
line-of-sight halo. If a line-of-sight halo has an effective mass
larger than the detection threshold for the subhalo, it is considered
detectable. However, in many cases, the line-of-sight halo at redshift
$z$ is not a good description of the image perturbation. As a result,
although one can always find a particular value of $m$ for a
line-of-sight halo that gives the smallest ${\chi}^2$ in the
fit of the image distortion generated by the subhalo, the two models
are not equivalent.

\begin{figure}
        \includegraphics[width=0.47\textwidth]{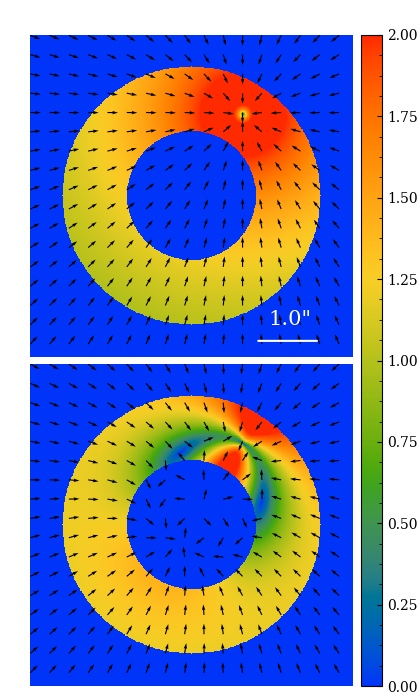}
        \caption{Comparison of the deflection angles caused by a
          $10^9$~M$_{\odot}$ NFW halo at lens plane ($z=0.5$) (upper
          panel) and a line-of-sight NFW halo in front of the lens plane
          ($z=0.2$) whose mass is $10^{8.67}$~M$_{\odot}$, as derived by
          fitting the deflection angle of the NFW halo at the main lens plane (lower
          panel). Both panels are derived by subtracting the total deflection angles of both main lens and perturber with the deflection generated only by the main lens. The colours show the amplitude of deflections (in units of $0.001$\arcsec). The arrows represent the direction of deflection
          angle vectors. In the lower panel, the asymmetric pattern of arrows at the centre of the main lens is subject to numerical noise because the deflection angle at the exact centre of an SIS is not well defined, where the profile's density is infinite and the density gradient (deflection angle) is not continuous. Since the perturber and the SIS are not on the same plane in the lower panel case, the inaccurate angles next to the SIS's centre are then not perfectly subtracted, which results in the asymmetric pattern of arrows in the centre. The arrows in other locations are reliable.}
        \label{fig:deflection_pattern}
\end{figure}

In Fig.~\ref{fig:deflection_pattern}, we show how the deflection angles produced by our lens are altered by the addition of a perturbing halo. The main lens is an singular isothermal sphere (SIS) located at
$z = 0.5$, at the centre of the image, and the source is at $z =
1.0$. The upper panel shows the deflections caused by a
$m_{200}=10^9$~M$_\odot$ NFW halo located on the main lens plane; the colour
indicates the amplitude of the deflections (in units of $0.001$\arcsec) and the arrows mark their
directions. In this case there are no non-linear multi-plane lensing effects, and so the change in the deflection angles due to a perturbing halo are just the deflection angles of the perturbing halo itself, which point towards the perturber's centre. In the lower panel we plot the change in the total deflection angles when a
$m_{200}=10^{8.67}$~M$_\odot$ NFW halo at $z = 0.2$ perturbs the lensing due to the main lens (subtracting the total deflection angle with the deflection caused only by the main lens). Because of multi-plane lensing effects \citep{Schneider1992, Fleury2021} the change in the deflection angles are no longer isotropic about the perturber centre \citep{Gilman2019, He2022}. Tracing from the observer backwards, the deflection of light rays by the perturber alters where those rays intersect the main lens plane, which in turn alters the deflection angles those rays receives from the main lens. It is clear that
the lensing effects of line-of-sight perturbers at different redshifts
cannot be reproduced by appropriately scaled subhaloes since the deflection patterns in the two cases
are completely different. Note that the mass of the line-of-sight perturber in the lower panel was chosen to best reproduce the deflection angle field in the annulus between 1.0\arcsec to 2.0\arcsec of the case in the top panel with a $10^9$~M$_\odot$ halo in the main lens plane (following Eq.~15 in D18), so according to Li17 and D18, the perturbers in the two panels have the same ``effective mass'', but clearly they have quite different effects.  

Degeneracies between the effects of the low-mass perturber and the main lens can also affect our estimation. As suggested in
Fig.~\ref{fig:deflection_pattern}, the deflection angles far from the
perturber's centre can be easily absorbed by slightly shifting and
stretching the main lens. Furthermore, the degeneracies between the
effects of perturbers and the main lens galaxy can be different at
different redshifts due to their distinct deflection patterns, making
the problem even more complicated.



In order to take into full account the complex effects discussed above, in this paper  we no longer compare deflection angles as Li17 and D18 did.  Instead, to quantify the  lensing effects of perturbers at different redshifts we directly fit image fluxes, a procedure that more closely reproduces what would happen on real data. We now define a new threshold for
detection through the log-likelihood\footnote{The log likelihoods are defined using a natural (base $\mathrm{e}$) logarithm.} improvement brought
about by including a perturber when fitting lensing images. Specifically, we first fit the mock
image with only a main lens and record the maximum log-likelihood value. We then fit the same image with a
model containing both a main lens and a perturbing halo and record
the log-likelihood of the best-fit model as well. If the log-likelihood
difference between the two fits is larger than a pre-established
threshold, we consider the perturber to be detectable.

In the tests we have carried out, the uncertainty in the modelling
comes exclusively from the statistical noise in the image data. The log-likelihood difference can be directly related to a significance level,
with a log-likelihood difference of 10 roughly corresponding to 4 $\sim$
$5\sigma$ significance. Note that in real observations, a threshold
based on the log-likelihood difference or the Bayesian evidence might not
be readily related to the true significance of a detection due to the
possible presence of various systematic effects in the data and 
analysis method \citep{Vegetti2012, Ritondale2019}.



\subsection{Number density of perturbers}


For each mock lens, we calculate the number of line-of-sight perturbing haloes within a radius of 3\,arcsec. The total number of line-of-sight perturbers with a mass in the range $[m_{\rm low}, 
m_{\rm high}]$ in a light cone corresponding to the $i^\mathrm{th}$ pixel can
be written as,

\begin{equation}
N_{\rm los} = \int_0^{z_s} \int_{m_{\rm low}}^{m_{\rm high}} \frac{\dif^{2} N}{\dif m \dif V} \frac{\dif V_{\rm sens}}{\dif z} \dif m \dif z \,, 
\end{equation}
where $\frac{d^2 N}{dm dV}\left(z\right)$ is the halo mass function at redshift $z$
\citep[e.g.][]{Sheth2001} and 
\begin{equation}
    \dif V_{\rm sens} = \Omega_{\rm sens}(m,z)\, \chi^2 \frac{\dif \chi}{\dif z} \dif z \,,
\end{equation}
where $\Omega_{\rm sens}(m,z)$ is the total solid angle corresponding to the areas on the redshift plane of $z$ that are sensitive to perturbers more massive than $m$.


For comparison, we also calculate the number of detectable subhaloes
for each mock image. For CDM, high-resolution N-body simulations have
shown that the mass function of subhaloes follows a power law
\citep{Springel2008}. Thus, the cumulative perturber density of
subhaloes in the mass range, $[m_1, m_2]$, in a host halo of mass, 
$m_{\rm 200}$, may be written as,
\begin{equation}
    \Sigma_{\rm sub,cdm}(m_1<m<m_2|m_{\rm 200})= \frac{\Sigma_0}{1-\alpha}\left( \left( \frac{m_2}{{\rm M}_\odot}\right)^{1-\alpha} -  \left( \frac{m_1}{{\rm M}_\odot}\right)^{1-\alpha} \right) \,, 
\end{equation}
where $\alpha=1.9$ \citep{Springel2008, Gao2012} and $\Sigma_0$ is a
normalization parameter that depends on $m_{\rm 200}$ and can be
determined from cosmological simulations. In this work, we estimate
this normalization using the same simulation \citep{Richings2021} that
we used to extract the density structure of subhaloes.


To obtain a sufficiently large
sample of subhaloes within the region of interest in order to
derive the subhalo mass function, we follow the same strategy as
before: we rotate the lens 10000 times, and only select those
subhaloes that fall on the region of interest in projection. We compute
the average number of subhaloes of mass $10^7\sim10^{11}$~M$_\odot$
for each projection. To obtain an estimate on the error, we repeat the same
procedure 200 times, each time resampling from all the 
subhaloes in the simulation before projecting along 10000 different lines-of-sight. We take the median value and 1$\sigma$ limits as our
estimate. Assuming $\alpha=1.9$, for a lens at $z = 0.22$, we derive
the normalization (from the snapshot at $z = 0.183$) to be
$(6.7\pm0.6)\times10^5\,{\rm arcsec}^{-2}$. For a lens at $z = 0.5$, we
derive the normalization (from the snapshot at $z = 0.503$) to be
$(2.3\pm0.2)\times10^6\,{\rm arcsec}^{-2}$.

To ensure that our result based on this one particular simulated  halo is not an outlier, we  compute the normalisation of the subhalo mass function from haloes in the  \textsc{Eagle} simulation in a similar way as above. Due to the poorer  resolution of \textsc{Eagle}, when calculating the normalisation we only count subhaloes of mass $10^9\sim10^{11}$~M$_\odot$. We select haloes in \textsc{Eagle} with mass within 0.1 dex of that of the main halo used in this study   and apply the same method to compute  $\Sigma_0$. For $z = 0.22$, there are 78 haloes and  $\Sigma_0=\left(4.3\pm1.5\right)\times10^5$~arcsec$^{-2}$, where the errors indicate the $1\sigma$ (34\%) scatter. For $z = 0.5$, there are 63 \textsc{Eagle} haloes and  $\Sigma_0=\left(1.8\pm0.9\right)\times10^6$~arcsec$^{-2}$. We see that the number of  \textsc{Eagle} subhaloes around halos of a given mass has a large scatter and  there can be differences of $2\sim3$ times within  \textsc{Eagle} itself.  At $z = 0.22$, the vaule of $\Sigma_0$ obtained for the halo analysed here is $1\sigma$ high compared to the distribution in \textsc{Eagle}, while at $z = 0.5$, the results are in even better  agreement. In conclusion,  the value of  $\Sigma_0$ derived in this work  is comparable to values for  \textsc{Eagle} haloes and our results based on one particular resimulation should be  representative.

\section{Results}\label{sec:results}

\begin{figure*}
        \centering
        \includegraphics[width=1.0\textwidth]{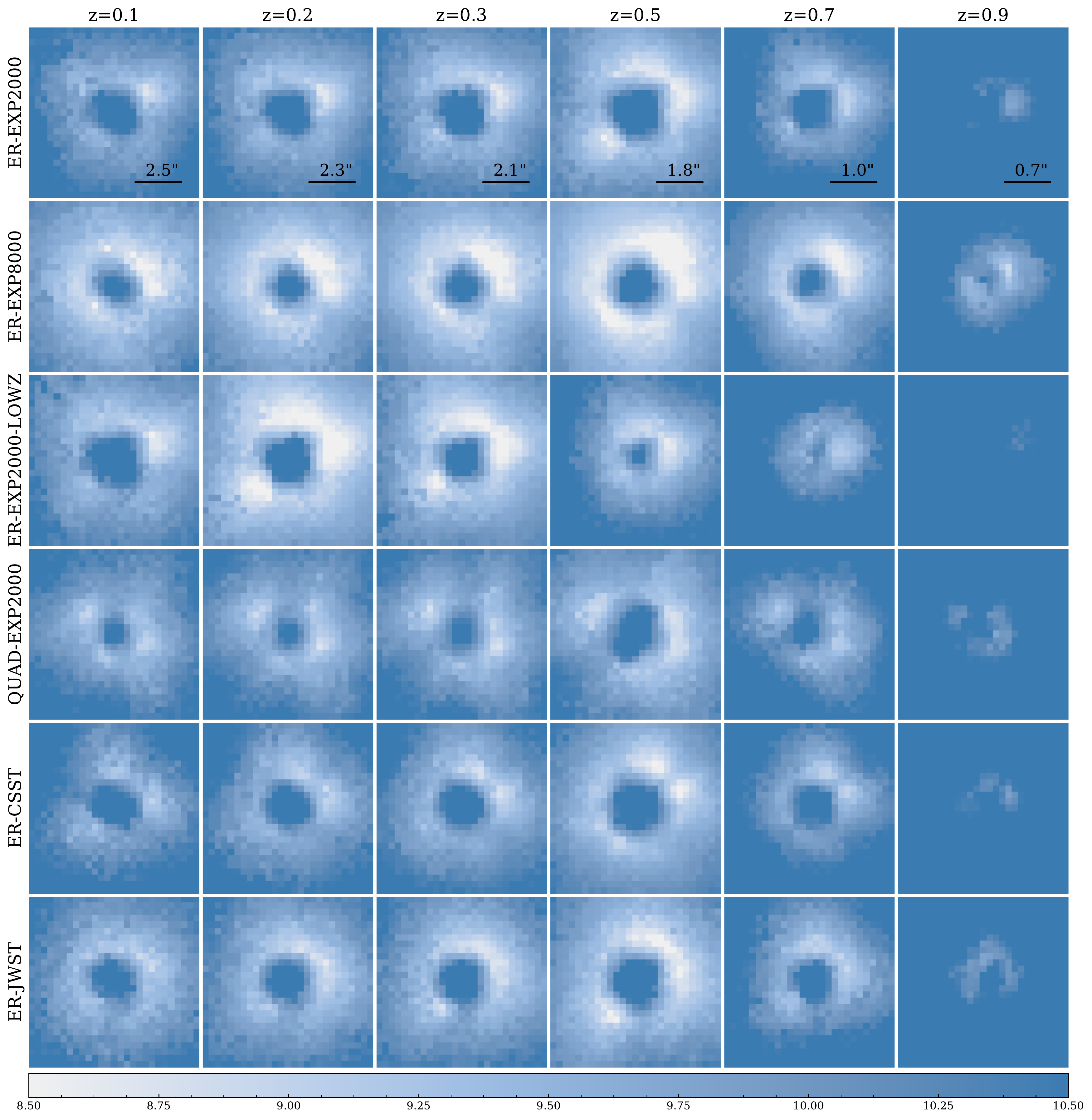}
        \caption{The sensitivity function of LOS perturbers. Each
          subpanel shows the detection limit for a perturber (a line-of-sight NFW halo) placed at a given redshift plane.  The
          colour bar gives the scale of $\log_{10}\left({m_{\rm th} / \mathrm{M}
          _{\odot}}\right)$. 6 panels on the same row show the sensitivity function for a mock lens at 6
          redshifts: 0.1, 0.2, 0.3, 0.5, 0.7, 0.9. For every column, the image size is marked by a scale bar in the top row. The label of each mock
          is given at the left of each row. In all cases the source
          is at $z=1$ and the lens is at $z=0.5$, except in the third
          row, where the lens is at $z=0.22$. The image sizes decrease with the redshit is because the angular size of regions of interests decreases with the redshift.}
        \label{fig:sensmap}
    \end{figure*}

In Fig.~\ref{fig:sensmap} we show the sensitivity function for our
mock lens images as a function of redshift. Each sub-panel displays a
map of $m_{\rm th}$ (see colour bar) for line-of-sight haloes placed
at a given redshift plane, for 5 redshifts. In all cases, the source
is at $z=1$ and the lens at $z=0.5$, except in the third row, where
the lens is at $z=0.22$. For all mock lenses, the threshold mass of a
detectable perturber is lowest near the lens redshift and raises
rapidly towards both higher and lower redshifts; however, the general
pattern of the sensitivity maps remains similar at each redshift.
This pattern varies considerably from one lensing system to
another. Visually, it appears similar to the corresponding pattern of
the lensing image. The value of $m_{\rm th}$ is lowest in the region
where the surface brightness is highest and highest in the region
where there is no light. Although the sensitivity maps for different
systems are quite different, the trend of their evolution with
redshift is similar.
     
The ability to detect low-mass dark haloes increases significantly
with the exposure time of the imaging. For our fiducial 2000s exposures, our
lensing systems are sensitive to line-of-sight perturbers of mass
$\sim 10^8$~M$_{\odot}$ only around the lens redshift. For 8000s,
however, perturbers of mass $\sim 10^8$~M$_{\odot}$ can be detected
over a much broader redshift range, from $z = 0.1$ to $z = 0.7$.  In the
bottom two rows of Fig.~\ref{fig:sensmap}, we show the sensitivity function
for the imaging quality achievable with the CSST and JWST, which have different resolutions. We find that with a lower image resolution, CSST lensing
images are still sensitive to perturbers of mass
$\sim 10^8$~M$_{\odot}$, although the overall sensitivity is somewhat
lower than with HST resolution. While for the JWST resolution imaging, the sensitivity is higher. 

\begin{figure}
        \centering
        \includegraphics[width=0.5\textwidth]{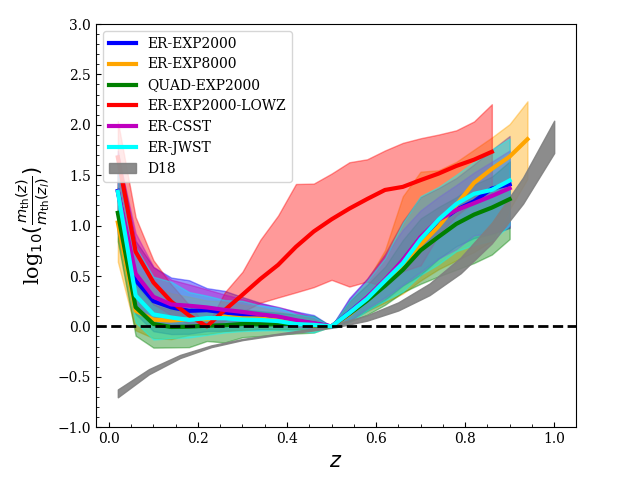}
        \caption{The mass threshold-redshift relation,
          $\log_{10}\left(\frac{m_{\rm th}(z)}{m_{\rm th}(z_l)}\right)$, as
          a function of redshift, $z$, for all of our mock settings.
          Four of them have a lens at $z=0.5$ and a source is at $z=1$, while one has a lens at $z=0.22$. The relation is
          calculated for each pixel on the image plane. The shaded
          regions enclose 70\% of the pixels.  For comparison, the
          relation and the scatter derived by D18 for our fiducial setting are shown in grey. The relations shown here are only for line-of-sight NFW perturbers and there is no subhalo (tNFW perturbers) involved in this comparison.}
        \label{fig:mass_redshift_all}
    \end{figure}

We now turn our attention to the all-important question of whether the
distortions to the Einstein rings are dominated by line-of-sight
perturbers or by subhaloes. We trace the position of each pixel on
the image plane at a series of redshifts and calculate the threshold
mass for detection, $m_{\rm th}\left(z\right)$. In
Fig.~\ref{fig:mass_redshift_all}, we plot the ratio,
$\log_{10}\left({m_{\rm th}(z)}/{m_{\rm th}(z_l)}\right)$, as a
function of redshift, $z$, for our different mock lensing systems. The
colour bands are the regions enclosing 70\% of the pixels, while the
means are shown as solid lines. For comparison, we also plot the
relation and scatter derived by D18 in grey. In D18 (as well as in
Li17), a line-of-sight halo at lower redshift is easier to detect than
a halo of the same mass at the lens redshift. Our new calculations
predict the different behaviour. For all configurations, the detection
threshold mass increases with $\Delta z= \left|z - \zl\right|$. For
lenses at redshift $\zl=0.5$, the detection threshold for
line-of-sight haloes at $z=0.1$ is $\sim 0.3$ dex higher than for
haloes at $\zl$.
    
\begin{figure}
        \centering
        \includegraphics[width=0.5\textwidth]{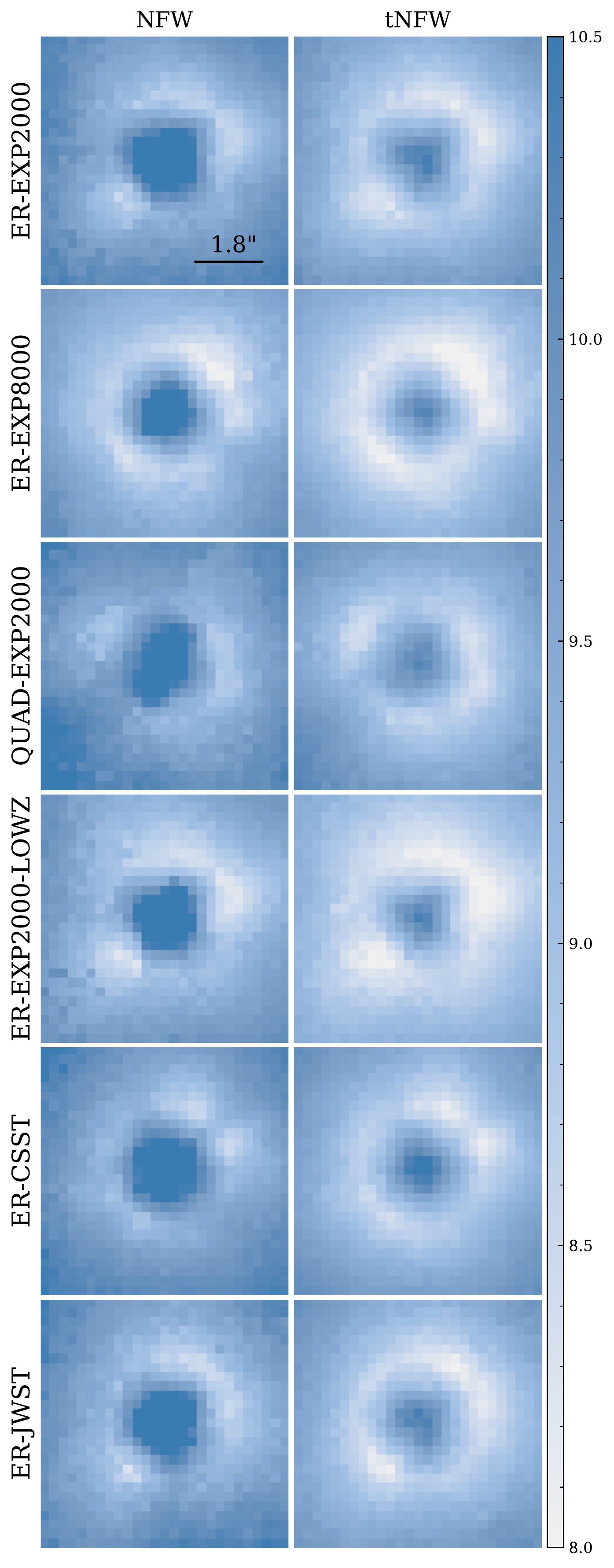}
        \caption{Sensitivity maps for NFW line-of-sight perturbers
          (left) and for tNFW subhalo perturbers (right). In
          all cases the source is at $z=1$ and the lens at $z=0.5$,
          except in the fourth row, where the lens is at $z=0.22$. All perturbers are on the same redshift plane of the main lens.}
        \label{fig:sens_sub}
    \end{figure}

To predict the number density of detectable subhaloes, we calculate
sensitivity maps for subhaloes with truncated NFW profiles, as
described in Sec.~\ref{sec:mock_lens}. In Fig.~\ref{fig:sens_sub}, we
compare sensitivity maps for subhaloes (right) to those for
line-of-sight haloes placed at the lens redshift of the same lensing
system (left). The maps on the left are the same as in the subpanels
of Fig.~\ref{fig:sensmap} at the corresponding (lens) redshift, while
the right panels show the detection limits for subhaloes. The mass of an NFW halo is defined to be $m_{\rm 200}$, while the mass of the
subhalo (tNFW) is defined as the total mass given by
Eq.~\ref{eq:mtot_tnfw}. As expected, the threshold mass for detecting
a subhalo is lower than that for detecting an NFW halo of the
same mass by about 0.5 dex, because a subhalo is much more compact than a halo of the
same mass. In the following calculations, we will
use the sensitivity maps for the tNFW haloes to estimate the number
of detectable subhalo perturbers

     %
    %
    


\begin{figure*}
        \centering
        \includegraphics[width=1.0\textwidth]{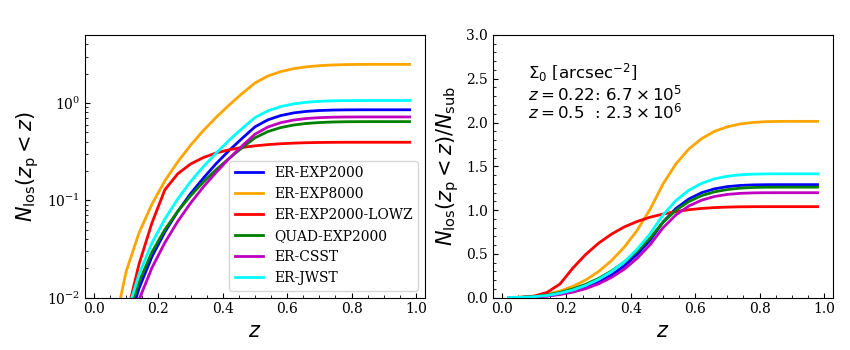}
        \caption{The cumulative number of detectable line-of-sight
          perturbers per lens for different lens configurations as a
          function of redshift (left). The number of line-of-sight
          haloes relative to the number of detectable subhaloes as a
          function of redshift (right). In all cases the source is
          at $z=1$ and the lens at $z=0.5$, except for the red line,
          where the lens is at $z=0.22$. The normalization of the subhalo mass function at two main lens redshifts is also listed on the right panel.} 
          
        \label{fig:NperLens}
    \end{figure*}

In the left panel of Fig.~\ref{fig:NperLens}, we show, as solid lines,
the expected cumulative number of detectable line-of-sight perturbers
derived, as a function of redshift; different colours correspond to
different systems. We can see that the number of detections rises
sharply around the lens redshift and then becomes flat. For the two
different mock lenses at $\zl = 0.5$, with a 2000s exposure (blue and green lines), 0.85 and 0.64 line-of-sight haloes can be detected per
lens. The differences between the two configurations are small, which may be due to the fact that the two mock
images have a similar number of high S/N pixels. For the lens
at $\zl=0.22$, the number of expected detections decreases to about
0.39 per lens (red line), which is $\sim 45\%$ of that of the
$\zl=0.5$ lenses. Fig.~\ref{fig:NperLens} also shows that a high S/N ratio (blue
line) helps reveal low-mass perturbers: if the exposure time increases
to 8000s, the number of detectable line-of-sight perturbers increases to 2.5 per
lens. We also see that by increasing the image resolution, the detectability of small perturbers increases. With CSST resolution only $~\sim$ 0.72 line-of-sight perturber can be detected per lens, but with a higher resolution as the JWST, the detectable number increases to 1.06 per lens.

On the right panel of Fig.~\ref{fig:NperLens}, we show the relative
importance of line-of-sight haloes and subhaloes. According to our calculation, for mock lenses at $\zl = 0.5$, the predicted number of detectable line-of-sight haloes is about 1.3 times the number of detectable subhaloes. For the low redshift mock, $\zl = 0.22$, the line-of-sight halo contribution is lower and close to that of subhaloes. For the high S/N mock, the relative importance of line-of-sight haloes increases, such that it becomes $\sim$ 2.0 times the number of detectable subhaloes.


    %
    %
    %

    
\section{Discussion and Conclusion}\label{sec:conclusion}

In this study we have revisited a key question relevant to the search
for low-mass haloes in strong lensing systems: what is the relative
contribution to the sensitivity function of line-of-sight (main)
haloes versus subhaloes in the lens. The main difference
between this and previous works is that, instead of fitting the
deflection angle map or an idealized image set, we have quantified the
expected number of line-of-sight perturbers by means of realistic
modelling carried out with the \textbf{\sc PyAutoLens} lensing package
on a set of mock lensing images with realistic levels of
noise. Contrary to previous work, we find that the lensing effect of a
line-of-sight perturber is largest if the perturber is located near
the lens redshift, and the strength of the signal decreases rapidly as
the redshift difference between the perturber and the lens plane
increases. Two reasons account for the difference: firstly, previous work assumed that the effects of a perturber at one redshift could be accounted for by a perturber at a different redshift, while Fig.~\ref{fig:deflection_pattern} demonstrates that the effects can be quite different because of multi-plane lensing effects; secondly, previous work did not take into account the degeneracy between the main lens and the small halo in the fitting process, whereby changes to the main lens model can absorb a significant part of the small perturber's lensing signal. Our calculation shows that the contribution from
line-of-sight haloes is still important, but does not dominate the
total number of detectable perturbers for most of our mock lenses as
was previously thought: previous studies overestimated the expected
total number of perturbers.
    
In a sense, our new results present an unwanted challenge for the
interpretation of future detections of low-mass haloes. Unlike the
line-of-sight field dark matter haloes of interest (whose masses are
below the minimum required to make a galaxy) which are unaffected by
baryons and thus retain their pristine structure, subhaloes are
changed by their environment, e.g. tidal striping and disruption. A detailed quantification of these processes needs understanding in detail the structure of the
galaxy, including its baryonic component. This requires full modelling
of galaxy formation such as that presented by \cite{Richings2020} for
a lens system of the kind in which we are interested for low-mass
halo and subhalo detection. 

In this work we have assumed, for simplicity, that the distribution of
line-of-sight haloes is not correlated with the lens host halo. In
reality, line-of-sight haloes are more strongly, and anisotropically
clustered around the region of the host halo than average
\citep{Richings2021}. In a recent paper, \citet{Lazar2021}
investigated the number density of line-of-sight perturbers in the
simulations from the FIRE\footnote{ http://fire.northwestern.edu} and
IllustrisTNG\footnote{https://www.tng-project.org} projects; they find
that the number of haloes correlated with the lens is about 35\%
larger than average, in agreement with the results of
\cite{Richings2021}.  Future work aimed at constraining the nature of
the dark matter from strong lensing data will need to take this sort
of correlation into account.
    
One caveat of our work is that we do not consider the scatter in the
mass-concentration relation. For a halo or subhalo of a given mass,
the higher the concentration, the higher the central density and
lensing signal. A recent study by \citet{Minor2021} shows
that this effect can introduce a bias of 3 for a subhalo of mass
$10^9$~M$_{\odot}$ and 6 for one of mass $10^{10}$~M$_{\odot}$. In a more
recent study, \citet{Amorisco2022} show that the scatter in the
mass-concentration relation boosts the detection of line-of-sight
perturbers and helps distinguish between CDM and WDM. A halo of mass less than $m_{\rm th}$ but of higher
than average concentration may still produce a strong enough lensing
signal to be detected and vice versa. When the halo or subhalo mass
function rises at the low mass end, the effect of scatter in the
mass-concentration relation can boost the number of detectable
perturbers significantly, helping distinguish different dark matter models. We remind our readers that the results discussed here are for analyses of resolved lensing systems where the sources are extended. Effects of mass-concentration relation and multi-plane lensing have been taken into account in previous similar studies on constraining low-mass perturbers' abundance in lensing systems with an unresolved source \citep{Gilman2019, Gilman2020}.

    



In this work, we also show that the ability to detect low-mass haloes increases with the exposure time of the image. For example, increasing the exposure time from 2000s to 8000s, increases the number of total detectable perturbers by a factor of 2. At face value, deeper imaging may seem not quite as efficient an observing strategy as observing more lenses. However, longer exposure times crucially increase the sensitivity to haloes of lower mass, which are important in constraining the identity of the dark matter.


\section*{Software Citations}

This work used the following software packages:
\begin{itemize}
\item 
\href{https://github.com/astropy/astropy}{\text{Astropy}} 
\citep{astropy1, astropy2}
\item
\href{https://github.com/dfm/corner.py}{\text{corner.py}}
\citep{corner}
\item
\href{https://github.com/joshspeagle/dynesty}{\text{dynesty}}
\citep{dynesty}
\item
\href{https://bitbucket.org/bdiemer/colossus/src/master/}{\text{Colossus}}
\citep{colossus}
\item
\href{https://github.com/steven-murray/hmf}{\text{hmf}}
\citep{Murray2013}
\item 
\href{https://github.com/matplotlib/matplotlib}{\text{matplotlib}} 
\citep{matplotlib}
\item 
\href{https://github.com/numpy/numpy}{\text{NumPy}} 
\citep{numpy}
\item
\href{https://github.com/rhayes777/PyAutoFit}{\text{PyAutoFit}}
\citep{pyautofit}
\item 
\href{https://github.com/Jammy2211/PyAutoLens}{\text{PyAutoLens}} 
\citep{Nightingale2015, Nightingale2018, pyautolens}
\item 
\href{https://www.python.org/}{\text{Python}} 
\citep{python}
\item 
\href{https://github.com/scipy/scipy}{\text{Scipy}}
\citep{scipy}
\end{itemize}

\section{Acknowledgements}

We thank the anonymous referee for insightful comments that helped us improve our paper. RL acknowledge the support of National Nature Science Foundation of China (Nos 11988101,11773032,12022306)), the science research grants from the China Manned Space Project (No CMS-CSST-2021-B01, CMS-CSST-2021-A01), the support from K.C.Wong Education Foundation. QH, AA, CSF and SMC acknowledge support from the European Research Council (ERC) through Advanced Investigator grant DMIDAS (GA
786910). We also acknowledge support from the STFC Consolidated Grant
ST/T000244/1. JWN and RJM acknowledge support from the UKSA through awards ST/V001582/1 and ST/T002565/1; RJM is also supported by the Royal Society. AR is supported by the ERC through Horizon2020 grant EWC (AMD-776247-6). NCA is supported by an STFC/UKRI Ernest Rutherford Fellowship, Project Reference: ST/S004998/1. This work used the DiRAC@Durham facility managed by the Institute for Computational Cosmology on behalf of the STFC DiRAC HPC
Facility (www.dirac.ac.uk). The equipment was funded by BEIS capital
funding via STFC capital grants ST/K00042X/1, ST/P002293/1,
ST/R002371/1 and ST/S002502/1, Durham University and STFC operations
grant ST/R000832/1. DiRAC is part of the National e-infrastructure.

\section*{Data availability}
The data underlying this article will be shared on reasonable request to the corresponding author.

\bibliographystyle{mnras}
\bibliography{ref}

\end{document}